\documentclass[twocolumn,superscriptaddress,amsfont,amssymb,amsmath, showpacs,balancelastpage, nofootinbib]{revtex4-1}
\usepackage{graphicx,color}
\usepackage[unicode=true,pdfusetitle,bookmarks=true, bookmarksnumbered=true,bookmarksopen=true, bookmarksopenlevel=1,
breaklinks=false,pdfborder={0 0 0},backref=false,colorlinks=true]{hyperref}
\hypersetup{citecolor=blue,filecolor=blue,linkcolor=blue,urlcolor=blue} 
\usepackage[usenames,dvipsnames]{xcolor}
\usepackage{amssymb,amsmath,mathtools,mathrsfs,enumitem} 
\usepackage[utf8]{inputenc}
\usepackage{tikz}
\usetikzlibrary{arrows,positioning,decorations.markings,decorations.pathmorphing,calc}
\usepackage{color,colortbl}
\usepackage{multirow,hhline,array}
\definecolor{hyperref}{RGB}{026,028,087}

\def\gsim{ \lower .75ex \hbox{$\sim$} \llap{\raise .27ex \hbox{$>$}} }
\def\lsim{ \lower .75ex \hbox{$\sim$} \llap{\raise .27ex \hbox{$<$}} }

\newcommand{\D}{{\rm d}}

\newcommand{\nin}{\noindent}
\newcommand{\MPl}{M_{\rm P}}
\newcommand{\Mpl}{M_{\rm P}}
\newcommand{\Oo}{{{\cal O}(1)}}

\newcommand{\comment}[1]{}

\newcommand{\aI}{\alpha_i}
\newcommand{\aB}{\alpha_B}
\newcommand{\aK}{\alpha_K}
\newcommand{\aM}{\alpha_M}
\newcommand{\aT}{\alpha_T}

\newcommand{\dof}{{\it dof}}
\newcommand{\dofs}{{\it dofs}}

\def\nn{\nonumber}

\def\ni{\noindent}
\newcommand{\km}{{\rm km}}

\definecolor{Gray}{gray}{0.9}
\definecolor{LightCyan}{rgb}{0.88,1,1}

\graphicspath{{./}}

\begin{document}

\hypersetup{pageanchor=false} 
\title{Scalar-tensor cosmologies without screening}

\author{Johannes Noller}
\affiliation{DAMTP, University of Cambridge, Wilberforce Road, Cambridge CB3 0WA, U.K.}
\affiliation{Institute for Theoretical Studies, ETH Z\"urich, Clausiusstrasse 47, 8092 Z\"urich, Switzerland}
\affiliation{Institute for Particle Physics and Astrophysics, ETH Z\"urich, 8093 Z\"urich, Switzerland}
\author{Luca Santoni}
\affiliation{Department of Physics, Center for Theoretical Physics, Columbia University, 538 West 120th Street, New York, NY 10027, USA}
\author{Enrico Trincherini}
\author{Leonardo G. Trombetta}
\affiliation{Scuola Normale Superiore, Piazza dei Cavalieri 7, 56126, Pisa, Italy}
\affiliation{INFN - Sezione di Pisa, 56100 Pisa, Italy}

\begin{abstract}
Scalar-tensor theories are frequently only consistent with fifth force constraints in the presence of a screening mechanism, namely in order to suppress an otherwise unacceptably large coupling between the scalar and ordinary matter. Here we investigate precisely which subsets of Horndeski theories do not give rise to and/or require such a screening mechanism.
We investigate these subsets in detail, deriving their form and discussing how they are restricted upon imposing additional bounds from the speed of gravitational waves, solar system tests and cosmological observables.
Finally, we also identify what subsets of scalar-tensor theories precisely recover the predictions of standard (linearised) $\Lambda${}CDM cosmologies in the quasi-static limit.
\end{abstract}

\date{\today}
\maketitle

\section{Introduction} \label{sec:intro}

Searches for new gravitational physics are naturally associated with looking for new gravitational degrees of freedom (dof). This follows from the fact that general relativity (GR) is the unique consistent (low-energy) theory of a massless spin 2 field and associates deviations from GR with the presence of fifth forces. This is especially true for theories of dynamical dark energy, where new gravitational \dofs{} have to be light and can therefore be the carriers of long-range forces. Yet the presence of such fifth forces is tightly constrained in our local environment, so proposed models for new gravitational physics need to ensure that their new \dofs{} remain sufficiently dormant on such scales in order to e.g.~pass solar system tests. Mechanisms that ensure consistency with local tests in this way go by the name of `screening mechanisms'---see \cite{Khoury:2010xi,Babichev:2013usa} for a review of several examples. 
The alternative to this, having new gravitational \dofs{} without screening (yet still consistent with fifth force constraints), can be realised as long as ordinary matter fluctuations do not couple to fluctuations of the new \dof. 
Considering such theories is especially pertinent in view of known inefficiencies of screening mechanisms---see e.g.~\cite{Bloomfield:2014zfa,Burrage:2019afs}.
However, while it is straightforward to mandate the absence of a matter-scalar coupling at the level of the initial Lagrangian and background solution, a coupling between the corresponding fluctuations is generically re-generated around non-trivial backgrounds. This severely restricts the space of viable scalar-tensor theories without a screening mechanism, with quintessence theories \cite{Wetterich:1987fm,Ratra:1987rm} being an obvious example. 
Here we ask what the space of consistent scalar-tensor theories without the need of a screening mechanism is beyond this, identifying and discussing some non-trivial theories that can pass observational constraints without the need for a screening mechanism on e.g.~solar system scales. 
\\

\nin {\bf Scalar-tensor theories}: We will be working in the minimal context of theories, where new gravitational physics is associated to a single scalar \dof{}. More specifically, we will be working within the scope of Horndeski gravity \cite{Horndeski:1974wa,Deffayet:2011gz}\footnote{For the equivalence between the formulations \cite{Horndeski:1974wa} and \cite{Deffayet:2011gz}, see \cite{Kobayashi:2011nu}.}, the most general Lorentz-invariant scalar-tensor theory with second order equations of motion. As such our starting point is the action
\begin{equation}
S=\int\mathrm{d}^{4}x\,\sqrt{-g}\left[\sum_{i=2}^{5}{\cal L}_{i}\,+\mathcal{L}_{\text{m}}[\Psi_i, g_{\mu\nu}]\right]\,,\label{eq:lagrangian}
\end{equation}
where the matter Lagrangian minimally couples the metric $g_{\mu\nu}$ to the matter fields $\Psi_i$ (we are in Jordan frame) and the scalar-tensor Lagrangians ${\cal L}_i$ satisfy
\begin{align}
{\cal L}_{2} & = \Lambda_2^4 \, G_2~, \quad\quad\quad\quad {\cal L}_{3} = \frac{\Lambda_2^4}{\Lambda_3^3} G_{3}\cdot[\Phi]\,, \nn \\
{\cal L}_{4}  & = \frac{\Lambda_2^8}{\Lambda_3^6} G_{4} R + \frac{\Lambda_2^4}{\Lambda_3^6} ~ G_{4,X} \left( [\Phi]^2-[\Phi^2] \right)\,,   \label{LH} \\
{\cal L}_{5} & = \frac{\Lambda_2^{8}}{\Lambda_3^9} G_{5}G_{\mu\nu}\Phi^{\mu\nu}-\frac{1}{6} \frac{\Lambda_2^4}{\Lambda_3^9} G_{5,X}([\Phi]^3 -3[\Phi][\Phi^2]+2[\Phi^3]), \nn
\end{align}
where we use the $(- \text{ + + +})$ signature convention, $\phi$ is a scalar field, $X = -\tfrac{1}{2}\nabla_\mu \phi \nabla^\mu \phi/\Lambda_2^4$ is the scalar kinetic term, $\Phi^{\mu}_{\;\; \nu} \equiv  \nabla^\mu \nabla_\nu\phi$, the $G_i$ are dimensionless functions of $\phi/\MPl$  and $X$, and $G_{i,\phi}$ and $G_{i,X}$ denote the partial derivatives of the $G_i$ (with respect to these dimensionless arguments). $\Mpl=(8\pi G)^{-1/2}$ is the reduced Planck mass, square brackets denote the trace, e.g.~$ [\Phi^2] \equiv \nabla^\mu \nabla_\nu \phi \nabla^\nu \nabla_\mu\phi$ and we have three mass scales: $\MPl$,  $\Lambda_2$ and $\Lambda_3$,  which the consistency of the  effective expansion requires to satisfy $\MPl \gg\Lambda_2\gg\Lambda_3$.\footnote{Note that the scale associated with shift-symmetry breaking operators can in principle be different from $\Mpl$, provided that it is parametrically larger than $\Lambda_2$. However, to avoid proliferation of scales throughout the text, we set it to $\Mpl$ in \eqref{LH}.} The scale $\Lambda_3$ represents the cutoff of the  theory and can be in principle as low as $ \Lambda_2^{4/3}\Mpl^{-1/3}$. 
As such, in cosmology, these mass scales are conventionally taken to satisfy $\Lambda_2^2 = \MPl H_0$ and $\Lambda_3^3 = \MPl H_0^2$, which then ensures that all interactions can give $\Oo$ contributions to the background evolution.
Also note that, in the case in which the functions $G_i$ depend only on $X$,  the robustness of the effective theory \eqref{LH} with  $\Lambda_3\sim\Lambda_2^{4/3}\Mpl^{-1/3}$  has been shown in \cite{Pirtskhalava:2015nla,Santoni:2018rrx} on the basis of a weakly broken galileon symmetry that protects the effective couplings against large quantum corrections. 
For a discussion of examples for which this protection against large quantum corrections is maintained even in the absence of shift symmetry, see \cite{Noller:2018eht,Heisenberg:2020cyi}.
\\

\section{The linear regime} \label{sec:lin}

\nin {\bf Cosmology and matter-scalar couplings}:  
In \eqref{eq:lagrangian},  we are assuming that  there is no quadratic mixing between the scalar \dof{} and the metric perturbations when expanded around a Minkowski spacetime ($\bar{g}_{\mu\nu}=\eta_{\mu\nu}$, $\bar \phi=0$).\footnote{Any quadratic mixing between $\phi$ and the metric perturbations on a Minkowski background can be always removed completely by suitable field redefinitions. These would induce however a direct coupling between the scalar \dof{} and the matter fields $\Psi_i$ on flat space. We are assuming for simplicity that this is  absent when the theory \eqref{eq:lagrangian} is considered on short-distance scales.} In particular,  this implies that, since the matter sector only  couples minimally to gravity in \eqref{eq:lagrangian}, around flat spacetimes a quadratic  mixing between  matter fields and  the scalar \dof{} is generated  exclusively at loop level through graviton exchange, corresponding therefore to effects that are suppressed in powers of $1/\Mpl$.
The situation changes when the theory \eqref{eq:lagrangian} is expanded around non-trivial backgrounds sourced by a $\bar\phi\neq0$. Indeed, in these cases, a generated quadratic mixing between $\delta g_{\mu\nu}\equiv g_{\mu\nu}-\bar{g}_{\mu\nu}$  and $\delta\phi\equiv \phi-\bar{\phi}$ is generically responsible for reintroducing a direct tree-level  $\delta\phi \cdot T$ coupling between the scalar \dof{} and the matter sector, after the quadratic Lagrangian for perturbations is diagonalized.  This is at the origin of a fifth force, with potentially $\mathcal{O}(1)$ effects, if the scale $\Lambda_3$ is chosen in such a way that the higher derivative operators in \eqref{eq:lagrangian} can provide $\mathcal{O}(1)$ contributions to the background evolution.
The standard way to reconcile the presence of a cosmologically relevant $\phi$ with solar system tests of gravity is based on screening mechanisms.  These crucially rely on scalar non-linearities becoming large near local sources, in such a way as to modify the dynamics and suppress the scalar potential compared to the Newtonian one. 
In the present work,  we instead consider another possibility, which, to the best of our knowledge, was not discussed before: we look for the subset of theories in \eqref{eq:lagrangian} which do not lead to a modified dynamics on short scales around local sources, without necessarily requiring large scalar non-linearities, i.e.~without relying on any screening mechanism.  As we will see, such \textit{unscreened} theories will correspond to very specific choices for the Horndeski functions \eqref{LH}.
\\

\nin {\bf Linear cosmology}: For our ansatz \eqref{LH}, the freedom in the dynamics of linearised perturbations around cosmological backgrounds can be captured by just four functions $\aI$ of time. Specifically these are \cite{Bellini:2014fua}
\begin{align}
\frac{M^2}{\Mpl^2} &= 2\left(G_4-2XG_{4,X}+XG_{5,\phi}-\frac{{\dot \phi}H}{\Lambda_3^3} XG_{5,X}\right) , \nonumber \\
\frac{M^2}{\Mpl^2} \aB &= - 2\frac{\dot{\phi}}{\Mpl H}\left(XG_{3,X}+G_{4,\phi}+2XG_{4,\phi X}\right) \nonumber \\ & 
+8X\left(G_{4,X}+2XG_{4,XX}-G_{5,\phi}-XG_{5,\phi X}\right) \nonumber \\ & 
+2\frac{\dot{\phi}H}{\Lambda_3^3}X\left(3G_{5,X}+2XG_{5,XX}\right) \nonumber , \\ 
\frac{M^2}{\Mpl^2} \aT &= 2X\left[2G_{4,X}-2G_{5,\phi}- \frac{\ddot{\phi}-\dot{\phi}H}{\Lambda_3^3}G_{5,X}\right] \,,
\label{alphadef}
\end{align}
where we also define $HM^2\aM \equiv \tfrac{\D}{\D t}M^2$ and we have omitted the fourth free function, $\aK$, since it will not be relevant for us (see \cite{Bellini:2014fua} for its precise form). In \eqref{alphadef}, all the quantities should be considered as computed on the background. The key parameters for us will be $\aM$, the `running' of the effective Planck mass $M_{\rm P}^{\rm eff} \equiv M$; $\aB$, the ``braiding'' that quantifies kinetic mixing between the metric and scalar perturbations \cite{Deffayet:2010qz}; and $\aT$, the tensor speed excess, related to the speed of GWs via $c_{GW}^2 = 1 + \aT$. 
\\

\nin{\bf The quasi-static limit}:
Of particular relevance for us will be the quasi-static approximation, which applies on small scales (in effect, on all but the largest cosmological scales) and amounts to assuming $|\dot{X}| \sim {H} |X|\ll |\partial_i X|$ for any gravitational (metric or Horndeski) perturbation $X$. It is therefore both a sub-horizon limit ($k^2/a^2 \gg H^2$) as well as a static approximation (in practice it amounts to setting $\dot X = 0$). 
Importantly it is also an excellent approximation for solar system scales, given that we are dealing with non-relativistic sources on small scales in this context.  
Working in Newtonian gauge, we can write
\begin{equation}\label{Def4Pert}
\D s^2=-\left(1+2\Phi\right)\D t^2+ a^2\left(1-2\Psi\right)\D x_i \D x^i,
\end{equation}
and focus on the effective Poisson equation as well as on the gravitational slip $\gamma$, which we can write as 
\begin{equation}
\begin{split}
\frac{k^2}{a^2}\Phi &=-4\pi G_{\rm eff} \hat\mu(a) \delta\rho_\text{m},  \\
\frac{\Psi}{\Phi} &= \gamma(a).
\end{split}
\label{mrs}
\end{equation}
Here $\delta\rho_\text{m}$ describes fluctuations in the matter density,\footnote{Technically this term is $\rho_\text{m}\Delta$, where $\rho_\text{m}$ is the background matter density and $\Delta$ is the corresponding comoving (gauge invariant) density contrast, but in the non-relativistic limit we recover $\rho_\text{m}\Delta \sim \delta\rho_\text{m}$ to high accuracy.} $\Phi$ and $\Psi$ (in the Newtonian gauge) capture the gauge-invariant Bardeen potentials, and $\hat\mu$ is the (time-dependent) free function/parameter that encodes deviations from the usual (general relativistic) Poisson equation, while $\gamma$ measures the presence of an effective gravitational anisotropic stress. These two functions fully describe all deviations from general relativity for Horndeski theories in the quasi-static limit. So from the four $\aI$ considered above, only two orthogonal free functions remain in this limit. $G_{\rm eff}$ is the effective Newton's constant, defined as $G_{\rm eff}=1/(8\pi M^2)$. In terms of the standard Newton's constant $G=1/(8\pi \MPl^2)$, we therefore have $G_{\rm eff} = G\Mpl^2 /M^2 $. GR predictions are recovered when $\hat \mu = 1$ and $M = \Mpl$.
When relating the above effective Poisson and slip equations to the underlying theory, we will find it useful to use the following shorthand \cite{Alonso:2016suf,Lagos:2017hdr,Noller:2020afd}\footnote{Note that, following CLASS \cite{Lesgourgues:2011re,Blas:2011rf} conventions and in comparison to the standard way of writing the Friedmann equations, we have re-scaled $p$ and $\rho$ by a factor of 3 here.}
\begin{equation}
\begin{split}
\beta_1 &\equiv - \frac{3(\rho_{\rm tot} + p_{\rm tot})}{H^2M^2} - 2\frac{\dot{H}}{H^2} + \frac{\tfrac{\D}{\D t}{(\aB H)}}{H^2},  \\
\beta_2 &\equiv \aB (1+\aT) + 2(\aM - \aT),  \\
\beta_3 &\equiv (1+\aT)\beta_1 + (1+\aM)\beta_2,  \\
\beta_4 &\equiv \aB (\aT - \aM) - \tfrac{1}{2}\aB^2(1 + \aT).
\end{split}
\label{betas}
\end{equation}  %
In terms of these parameters we can find the following expressions for the quasi-static parameters $\hat\mu$ and $\gamma$,
\begin{equation}
\begin{split}
\hat\mu &= \frac{2\beta_3}{2\beta_1+\beta_2(2-\aB)},  \\
\gamma &= %
\frac{\beta_1+\beta_2}{\beta_3}.
\end{split}
\label{mugamma}
\end{equation}
In addition, in order to connect the quasi-static approximation to  lensing observables, it is often convenient to introduce  the following combination,
\begin{equation}
\hat\Sigma = \frac{1}{2}(1+\gamma) \hat\mu
=  \frac{\beta_1+\beta_2+\beta_3}{2\beta_1+\beta_2(2-\aB)} \, ,
\label{Sigmalens}
\end{equation}
which in GR reduces to $\hat\Sigma=1$, i.e.~$\hat\Sigma$ probes modifications to the effective lensing potential.\footnote{In the above we are introducing hatted variables $\hat\mu$ and $\hat\Sigma$, which are re-scaled versions of the standard $\mu$ and $\Sigma$ variables. This re-scaling amounts to a multiplication by $G_{\rm eff}/G$ and removes any explicit dependence of $\hat\mu$ and $\hat\Sigma$ on the absolute value of $M/\MPl$. This is useful, because a constant $M \neq \MPl$ can always be re-absorbed into pressure and density terms cosmologically (since these are only probed gravitationally), so while the time-variation of $M$ is observable here, its absolute value is not.} 
\\

\ni{\bf Linking cosmological and local scales}:
Having obtained an explicit form of the effective Poisson equation that relates the Newtonian potential to matter perturbations in terms of the $\aI$ parameters, we can now use this equation to link with observational constraints related to the movement of test masses in a gravitational potential $\Phi$. In practice, we will use the bounds from solar systems experiments to put constraints on the $\aI$ parameters entering the Poisson equation \eqref{mrs} and we will identify the Horndeski subclass of theories that pass the experimental tests without the need for screening.
However, before getting there, we would first like to elaborate more on some aspects underlying  this logic. When we claim to combine the cosmological Poisson equation with the bounds from solar system experiments, like Lunar Laser Ranging (LLR), we are tacitly taking for granted that, in the absence of screening, any modification to the gravitational potential on cosmological scales, encoded in  the effective Poisson equation, e.g.~a time dependence in  $G_{\rm eff}(t)$,   survives down to Earth-Moon distances in such a way that it is meaningful to translate LLR bounds into constraints on the $\aI$. This statement is proven to be pretty robust  in the context of shift-symmetric scalar-tensor theories, where the shift symmetry guarantees that the time-dependence in $\bar\phi(t)$ is inherited by the modified background solution on short scales \cite{Babichev:2011iz}. In this case, bounds from Earth-Moon experiments can be fairly used to constrain the form of the theory \eqref{eq:lagrangian}.  On the other hand,  in more general situations, even if it is still generically true, the picture is less clear and the answer might be model dependent.
In the following, we will ignore this caveat---see e.g.~\cite{Burrage:2020jkj} for a discussion about this point and also \cite{Belgacem:2018wtb} for a counterexample in the context of nonlocal theories of gravity. 
\\

\section{No screening} \label{sec:no-screen}
In the present section, after briefly reviewing the definition of Vainshtein radius, we will derive the condition for the absence of an induced $\delta\phi \cdot T$ coupling on the cosmological background. This corresponds to having neither screening around local sources nor fifth force between them. In particular, we shall see what this implies at the level of the theory \eqref{eq:lagrangian} and discuss the constraints from solar system experiments.
\\

\nin {\bf The Vainshtein radius}: 
Consider the induced quadratic mixing   between the scalar and ordinary matter \dofs, $(\delta/\MPl) \cdot \delta\phi \cdot T$, where we are denoting with $\delta$   the dimensionless coupling `constant'. This coupling being small amounts to requiring $\delta \ll 1$. The Vainshtein radius $r_V$ for Galileon-type interactions (the subset of \eqref{LH} with Galilean $\phi \to \phi + c+ b_\mu x^\mu$ symmetry \cite{Nicolis:2008in}) then satisfies \cite{Babichev:2013usa} 
\begin{align} \label{rV}
r_V = \frac{1}{\Lambda_3}\left(\frac{M_\star\delta}{8\pi \MPl}\right)^{\frac{1}{3}}.
\end{align}
Here $M_\star$ refers to the mass of the source (e.g.~the Earth or Sun). Within this radius classical non-linearities in the solution dominate and screen the scalar field $\phi$.  
Now, for $\Lambda_3^3 = H_0^2 \MPl$ and $\delta \sim 1$, the Vainshtein radius for the Sun is $0.1 \; {\rm kpc} \sim 10^{15} \; \km$, i.e.~much larger than the size of the solar system $\sim 10^{9} \; \km$.
If $\delta = 0$, then of course there is no coupling between $\delta\phi$ and $T$, so the Vainshtein radius vanishes and no such screening is present. How small does $\delta$ have to be for there to be no screening on local solar system scales and what does this imply for the underlying theory such as \eqref{LH}? Since $r_V$ scales with $\delta^{1/3}$, the absence of screening in the solar system in the presence of the higher derivative interactions in \eqref{LH} then requires 
\begin{align} \label{deltaEq}
\delta \ll 10^{-18}. 
\end{align}
Note that, while the precise numerical bound here will not matter for us, it depends on exactly how small we require the Vainshtein radius to be, i.e.~what the smallest scale mandated to be unaffected by screening effects is. Eq.~\eqref{deltaEq} is a conservative estimate derived via taking the size of the solar system as the relevant scale, but when e.g.~explicitly requiring the Vainshtein radius to be below Earth-Moon distance scales ($\sim 10^5 \, \km$) the bound is strengthened to $\delta \ll 10^{-30}$.   
\\

\nin {\bf Cosmological couplings}: We now wish to explicitly identify the scalar-matter couplings generated by cosmological backgrounds, i.e.~to establish the precise $\delta$ induced by such backgrounds. To do so, we focus on the quasi-static limit discussed above and consider the equation of motion for perturbations of the Horndeski scalar field $\delta\phi$. More specifically, we take the action describing linear perturbations around a cosmological background in Newtonian gauge (see appendix \ref{app:qa} for details) and derive the equations of motion for $\Phi, \Psi$ and $\delta\phi$. Specialising to the quasi-static limit,\footnote{In practice this amounts to setting all time derivatives of $\Phi, \Psi$ and $\delta\phi$ to zero, as well as any $k^2$-independent contributions. Note that no terms involving matter fluctuations are dropped, as our quasi-static approximation only applies to gravitational \dofs{}.} we can then use the equations of motion for $\Phi$ and $\Psi$ to solve for these fields, ending up with the following equation of motion for $\delta\phi$
\begin{align} \label{QSA_scalar}
k^2 \delta\phi = \beta_2 \cdot \frac{a^2 \dot{\bar{\phi}} }{2 H M^2 {\cal D} c_s^2} \cdot \delta\rho_m.
\end{align}
Here ${\cal D}c_s^2 = \beta_1 + \beta_2 + \beta_4$, where ${\cal D}$ and $c_s^2$ are both positive, when requiring the absence of ghost and gradient instabilities, respectively (where we also assume the absence of infinite strong coupling, which would be a consequence of either of these variables being zero). 
The key observation that then follows from \eqref{QSA_scalar}, is that $\delta \sim \beta_2$. So the scalar is sourced by matter on quasi-static scales, {\it iff} $\beta_2 \neq 0$, where we assume a cosmological background with positive $H$ and non-trivial background $\dot{\phi}$. Conversely, fully maintaining the decoupling between matter and scalar fluctuations at tree level around cosmological backgrounds to the level required by \eqref{deltaEq} then requires\footnote{Note that, for the precise numerical bound, we are implicitly assuming that $\dot{\bar\phi} \sim M H$ and ${\cal D}c_s^2 \sim {\cal O}(1)$ here. Otherwise that precise numerical bound for $\beta_2$ changes. But regardless of its precise value, the bound generically renders $\beta_2$ negligibly small for observational purposes.}
\begin{align} \label{b2condi}
\beta_2 \equiv \aB (1+\aT) + 2(\aM - \aT) \ll 10^{-18}.
\end{align} 
To all intents and purposes we will therefore set $\beta_2 = 0$ in what follows. 
From \eqref{mugamma} we can see that this also implies
\begin{align}
\hat\mu &= \frac{1}{\gamma} = 1 + \aT, 
&\hat\Sigma &= 1+\frac{\aT}{2}.
\label{mugamma2}
\end{align}
When requiring the absence of an induced matter-scalar coupling in the quasi-static regime, deviations from standard general relativistic behaviour (at the level of linear perturbations) are therefore solely controlled by $\aT$.
\\

\nin {\bf The speed of gravitational waves}: Before jumping into deriving   constraints from solar system tests, %
recall that the bounds on the measured speed of gravitational waves from GW170817 \cite{TheLIGOScientific:2017qsa} and GRB170817A \cite{Goldstein:2017mmi} impose $\aT \ll 10^{-15}$ \cite{Monitor:2017mdv} (also see \cite{Creminelli:2017sry,Sakstein:2017xjx,Ezquiaga:2017ekz,Baker:2017hug} and references therein for dark energy-related consequences of this observation), as long as the cosmological theory in question is still (sufficiently) applicable at the energy scales measured by GW170817 (see \cite{deRham:2018red} for a related discussion). Assuming that this is indeed the case, we can impose $\aT = 0$ to all cosmological intents and purposes.  
From \eqref{mugamma2}, we already know that imposing this requirement in addition to setting $\beta_2 = 0$ eliminates linear cosmological deviations from GR, except for possibly on the very largest scales, as $\hat \mu = 1 = \gamma$ now.
We will come back to this point below, but for now notice that, upon imposing $\aT = 0$, \eqref{b2condi} implies that\footnote{This is in agreement with the luminal gravitational wave and Horndeski limit of the scalar equation of motion derived e.g.~in \cite{Crisostomi:2019yfo}. Note that there is a conventional difference between the definition of $\aB$ employed here, which follows \cite{Bellini:2014fua}, and that of \cite{Crisostomi:2019yfo}, amounting to an overall factor of $-2$.}
\begin{align}
\aB = -2 \aM.
\end{align}
Mapping these two requirements ($\aT = 0$ and $\aB=-2\aM$) onto the Horndeski $G_i$ functions via \eqref{alphadef}, we find the following residual set of theories\footnote{We are using that a $G_5$ that depends only on $\phi$ can always be reabsorbed through a redefinition of the other Horndeski functions $G_2$, $G_3$ and $G_4$ following an integration-by-parts.}
\begin{align}
S &=\int\mathrm{d}^{4}x\,\sqrt{-g} \Big[ \Lambda_2^4 \, G_2(\phi,X) + \frac{\Lambda_2^4}{\Lambda_3^3} f'(\phi) \ln{X} [\Phi] \nn \\
&+  \MPl^2 f(\phi) R   \Big].
\label{ns_theory}
\end{align}
In other words, we have
\begin{align}
G_3 &= f'(\phi) \ln X, 
&G_4 &= f(\phi), &G_5 &= 0,
\end{align}
where $f(\phi)$ {has no term linear in $\phi$ (otherwise this would induce kinetic mixing and hence, upon diagonalisation, an effective scalar-matter coupling already around flat space)}, but is otherwise a free function of $\phi$. To recap, \eqref{ns_theory} is then the subset of Horndeski theories that do not induce a significant matter-scalar coupling on cosmological scales, while also yielding luminally propagating gravitational waves. 
{Note that I) the class of theories \eqref{ns_theory} is still richer than k-essence theories, II) requiring the absence of gradient and ghost stabilities (for a related recent discussion of `emergent' ghosts see \cite{Babichev:2020tct}), i.e. ${\cal D} > 0$ and $c_s^2 > 0$ as discussed above, places constraints on the two free functions in \eqref{ns_theory}. ${\cal D}$ depends on both $G_2$ and $f(\phi)$ and can straightforwardly be made positive with a judicious choice of $G_2$ (we refer to \cite{Bellini:2014fua} for the full expression of ${\cal D}$ in terms of the $G_i$). For $c_s^2$, on a $\Lambda{}$CDM background, we here find ${\cal D}c_s^2 = \tfrac{d}{dt}(\alpha_B H)/H^2$, where $\tfrac{M^2}{M_P^2}\alpha_B = -\tfrac{2}{H M_P} \tfrac{d}{dt}f(\phi)$. So the absence of gradient instabilities places a constraint on the form of $f(\phi)$, and III) an unavoidable feature in these theories is the nonanalytic $X$-dependence of the $G_i$ functions, which prevents them from having healthy perturbations around an $X=0$ background, i.e.~flat space. We will come back to this point later.} \\

\nin {\bf Solar system tests}:
Solar system tests constrain both the effective Poisson equation as well as the gravitational slip in \eqref{mugamma} and \eqref{Sigmalens}. The tests that we will consider here, which are relevant over distance scales in the domain of  validity of the effective theory \eqref{eq:lagrangian}, are lunar laser ranging (LLR) and light deflection experiments.\footnote{Observational  bounds in this context are frequently expressed in the language of the parametrized post-Newtonian (PPN) formalism \cite{Will:1993hxu}. While we will map these constraints directly onto cosmologically relevant partameters, for an explicit mapping between PPN parameters and the (cosmological and quasi-static) $\hat\mu$ and $\gamma$ parameters, we refer the reader to \cite{Clifton:2018cef}.
} 

Some of the tightest bounds come from Shapiro time delay measurements, in particular from observations of the frequency shift of radio photons from the Cassini spacecraft \cite{Bertotti:2003rm,Will:2005va,Hofmann:2018myc}. This can be re-cast as a bound on the metric potentials (on solar system scales) \cite{Bertschinger:2011kk}\footnote{Note that this bound interestingly does not involve the absolute value of $G_{\rm eff}$. This is analogous to the cosmological case, where any $M \neq \MPl$ could always be absorbed into cosmological densities and pressures. There, this was the case because we only have gravitational measurements of cosmological fluids. In the solar system we also only know about the mass of lenses (i.e.~the sun, in the case of Cassini constraints) gravitationally, so the absolute value of any $M \neq \MPl$ can also be absorbed into the mass itself. Note that earthbound measurements of $G$, if considered to be within the regime of validity of the EFT, can in principle break this degeneracy and that the time variation of $M$, i.e.~$\aM$, is observable as before (see the LLR discussion below).}
\begin{align}
\left|\frac{\Psi-\Phi}{\Phi}\right| \lesssim 10^{-5} \quad\quad \Rightarrow \quad\quad \left| \gamma - 1 \right| \lesssim  10^{-5}.
\end{align}
Comparing this with \eqref{mugamma2} yields $\vert\aT\vert\lesssim 10^{-5}$. 
As before, with $\hat\mu = 1 = \gamma$ now at the $10^{-5}$ level, and since the quasi-static approximation is expected to apply to the vast majority of cosmological scales, this analysis already tells us that the theory \eqref{ns_theory} should closely resemble $\Lambda{}$CDM predictions for cosmological observables, with the only significant deviations expected on very large scales. 
In fact, since current cosmological constraints for the $\aI$ are of order one (see e.g.~\cite{mcmc,Frusciante:2019xia} and references therein) and near-future constraints are expected to be of order $10^{-1}$ \cite{Alonso:2016suf}, this means we effectively have $\aT = 0$ for all current and near-future cosmological applications, regardless of whether we are also imposing the gravitational wave speed bound from GW170817 discussed above (note that the Cassini bound probes energy scales several orders of magnitude below those probed by LIGO). We have therefore again reduced our theory to \eqref{ns_theory}, with $\aB = -2\aM$.

Secondly, LLR can be used to put a constraint on the time variation of the  coupling appearing in the effective Poisson equation \eqref{mrs}. 
Here, one can distinguish two contributions that potentially lead to a modified motion of test masses in an external gravitational field (with respect to GR predictions). The first is given by $G_{\rm eff}$, which basically encodes variations in the Planck mass in the Einstein-Hilbert action. The second is $\hat \mu$, which can instead be thought of as resulting from a modified energy-momentum tensor in the Einstein equations. In general, LLR will therefore probe the time-dependence induced by these two contributions. However, given the Cassini and/or GW170817 constraints discussed above, here we now effectively have $\hat \mu = 1$ (up to at most one part in $10^5$), so the only potentially important remaining contribution is the one from $G_{\rm eff}$. Experimental bounds from lunar laser ranging LLR---see e.g.~\cite{Uzan:2002vq,Williams:2004qba,Merkowitz:2010kka} and \cite{Hofmann:2018myc} for the most recent result---then impose
\begin{equation}
\left|\frac{\frac{\D}{\D t} G_{\rm eff}}{G_{\rm eff}}\right| = \left|H \aM\right| \lesssim 10^{-3} H_0,
\label{caM}
\end{equation}
so the relative variation of  $G_{\rm eff}$ per unit Hubble time is smaller than $\lesssim 10^{-3}$ and hence $|\aM| \lesssim 10^{-3}$.\footnote{Note that in general $\frac{\D}{\D t} (G_{\rm eff} \hat\mu)/(G_{\rm eff} \hat\mu) = \dot G_{\rm eff}/G_{\rm eff} + \dot {\hat\mu}/\hat\mu$ is the parameter combination that is constrained by this bound and hence at most ${\cal O}(10^{-3})$, so in the presence of a non-trivial $\hat\mu$, the time-variation of $\hat\mu$ also needs to be taken into account carefully.
For example, taking $\hat\mu$ from \eqref{mugamma2} without imposing additional constraints on $\aT$, the LLR constraint becomes
\begin{equation}
\frac{\frac{\D}{\D t}( G_{\rm eff} \hat\mu)}{G_{\rm eff} \hat\mu} =\frac{ \dot G_{\rm eff}}{G_{\rm eff}} +\frac{ \dot {\hat\mu}}{\hat\mu}
= -H\aM + \frac{ \dot{\aT}}{1+\aT}  \, .
\label{eq00}
\end{equation}
In the absence of an orthogonal bound on $\aT$, one is then in principle able to `tune' the time-dependence of $\aT$ to compensate the time-dependence in $G_{\rm eff}$ and pass the LLR test, while allowing for non-suppressed $\aM$ and $\aT$. Given the Cassini/GW170817 bounds (and assuming that $\dot\aT \sim H \aT$ is a good order-of-magnitude estimate), this possibility is ruled out, however.
}

Combining Cassini and LLR bounds, we therefore find the following resulting constraints from solar system measurements alone
\begin{align}
&\left|\aT\right| \lesssim 10^{-5}, &\left|\aB\right| &= 2\left|\aM\right| \lesssim 10^{-3}.
\label{bound12}
\end{align}
{If the condition $\left|\aT\right| \lesssim 10^{-5}$ is used, together with $\aB = -2\aM$, to cast the theory in the form \eqref{ns_theory}, then the second bound in \eqref{bound12} translates into
\begin{equation}
2\sqrt{X} \frac{f'(\phi)}{f(\phi)}  \lesssim 10^{-3}. 
\end{equation}
}
Given the precision of cosmological constraints on the $\aI$ discussed above, future improvements on the already tight $\aI$ bounds for these theories are therefore most likely to come from increasingly accurate solar system measurements themselves. As far as linear cosmology is concerned, this then essentially renders all Horndeski theories equivalent to k-essence and quintessence, i.e.~pure $G_2$ theories \cite{ArmendarizPicon:1999rj,Garriga:1999vw,ArmendarizPicon:2000dh,ArmendarizPicon:2000ah}. Notice that, even if the effects of higher derivative operators in these subclasses of theories are highly suppressed on cosmological and solar  system distances, they could still affect the physics of systems on much shorter scales and leave imprints on different types of observables -- see \cite{Noller:2019chl} for an example. In addition, there might also be other operators, on top of those in \eqref{ns_theory}, that, even if negligible on the cosmological background,  give rise to sizeable effects on smaller scales \cite{Noller:2019chl}.

\section{Recovering GR predictions}

In the previous section we answered the question of which sets of theories {\it do not need} a screening mechanism, since the extra gravitational \dof{} is effectively not sourced by matter. However, one could be interested in principle in a slightly different, but  closely related, question: which subset of theories  recovers linearised GR predictions, irrespective of the existence of an induced direct coupling to matter? In other words,  to what extent is it possible to find theories where the effects of the extra \dof{} are  suppressed, as far as the linearised dynamics of perturbations is concerned, regardless of the existence of a screening mechanism close to matter sources?\footnote{Clearly, such a kind of question is relevant, if there is no screening mechanism in the first place. However, there are other situations, like the one discussed in \cite{Hui:2012jb} of a long wavelength scalar field acting on  systems with much shorter Vainshtein radius, in which, even if present, one could still see some effects due to the different responses of the objects within the system. In the spirit of our analysis, the non-observation of such effects would not necessarily translate into a bound on the coupling $\delta$ \cite{Hui:2012jb,Sakstein:2017bws}, but, as we discuss in the present section, it could also be interpreted as a constraint on the specific form of the higher-derivative operators in the scalar-tensor theory \eqref{eq:lagrangian}.}
Here we will try to answer this question at the level of  Eqs.~\eqref{mugamma} by  explicitly looking for the subset of Horndeski theories that leave the linearised  GR dynamics of test masses unchanged,  and that are therefore compatible with current bounds from linear cosmology and solar system tests by design, regardless of whether the additional gravitational \dof{} is sourced by matter or not. {Note that this question is also closely related to the analysis carried out in \cite{Babichev:2016kdt}, where (among other findings) a subset of shift-symmetric Horndeski theories that display cosmological self-tuning and possess a local metric around a spherical body which is indistinguishable from GR was considered.}
\\

\nin{\bf Linearised dynamics}: In the following, we  keep the discussion generic, without committing to any specific example or making any assumption about the size and the typical scales of the fields' non-linearities. In this sense, we  generically require $\gamma \simeq 1$ and $G_\text{eff}\hat{\mu}\simeq G$, without  quantifying the tolerable range of values for    possible deviations from GR. Clearly, in realistic situations, these equalities, as well as the results below,  will be accompanied by appropriate numerical bounds.
Let us start requiring the absence of gravitational slip, $\vert\gamma-1\vert\simeq0$. Looking at Eq.~\eqref{mugamma}, this amounts to 
\begin{align} \label{slipCon}
\aT \beta_1 + \aM \beta_2 = 0 \, .
\end{align}
On the other hand, simultaneously imposing $\hat \mu = 1$ (or $\hat\Sigma=1$) adds the following condition
\begin{align} \label{muCon}
\aB \beta_2 = 0.
\end{align}
Let us focus  on the gravitational slip requirement for the time being. Eq.~\eqref{slipCon} is in principle a complicated non-linear equation for the $\alpha$-parameters and, in turn, for the Horndeski functions $G_i(\phi,X)$. However, imposing that it remains valid at any time and upon deformations of the $\Lambda$CDM background leads to the following three  possibilities: 
\begin{enumerate}[label=\textit{\roman*})]
\item $\aT=0=\beta_2$,
\item $\aT=0=\aM$, 
\item $\aM=0=\aB$.
\end{enumerate}
Note that case \textit{i)} corresponds to the setup discussed in the previous section, while the others may still allow for some residual $\mathcal{O}(1)$ coupling between the scalar and matter through a non-zero $\beta_2$.
Requiring in addition Eq.~\eqref{muCon} and  {$G_{\rm eff} \simeq  G$},\footnote{{The latter follows from requiring $G_\text{eff}\hat{\mu}\simeq G$ and $\hat \mu = 1$ (or $\hat\Sigma=1$).}  } irrespective of the particular background cosmological evolution,  cases \textit{i)} and \textit{ii)} become degenerate with $\aT=\aM=\aB=0$. 
Case \textit{iii)} is instead special, because after requiring $\aM=0=\aB$ the conditions \eqref{muCon} and $G_{\rm eff}\simeq \text{constant}$ are automatically fulfilled, as is $\gamma \simeq 1$.
In this case, all GR tests related to $\hat\mu$, $\gamma$ and $G_{\rm eff}$ are therefore passed by design (recall that $G_{\rm eff} = G$, since $\aM = 0$ here), but we retain $\aT$ as a free parameter. Finally, note that these cases also correspond to some of the `limited modified gravity' scenarios considered in \cite{Linder:2020xza}.
\\

\nin {\bf Horndeski subsets}: Remaining agnostic about the existence of screening, from the previous paragraph we conclude that, among the three possible options presented above that correspond to an unchanged GR dynamics at linear level, only case \textit{iii)} is potentially non-trivial.\footnote{Again, by `non-trivial' we mean that the covariant  Lagrangians belonging to the subset of theories  that are compatible with  such dynamics are not just equivalent to quintessence or k-essence type of models, as far as cosmological observations are concerned.} 
Thus, let us focus on case \textit{iii)} here. Mapping the requirement $\aM=0=\aB$ onto the Horndeski $G_i$ functions via \eqref{alphadef}, we now find the following residual set of theories:
\begin{align}
S &=\int\mathrm{d}^{4}x\,\sqrt{-g} \Big[ \Lambda_2^4 \, G_2(\phi,X) + 4 \frac{\Lambda_2^4}{\Lambda_3^3}\cdot f'(\phi) \sqrt{X} \cdot[\Phi] \nn \\
&+  \frac{\Lambda_2^8}{\Lambda_3^6} (c+f(\phi) \sqrt{X}) R + \frac{\Lambda_2^4}{\Lambda_3^6} \cdot \frac{f(\phi)}{2\sqrt{X}} \left( [\Phi]^2-[\Phi^2] \right)    \Big].
\label{aT_theory}
\end{align}
In other words, we have
\begin{align}
G_3 &= 4 f'(\phi) \sqrt{X}, 
&G_4 &= c + f(\phi) \sqrt{X}, &G_5 &= 0.
\end{align}
where $c$ is a constant and $f(\phi)$ is a free function of $\phi$.
In this case, setting $\aM=0=\aB$ is enough to recover the standard Poisson equation with $G_{\rm eff}=\text{constant}$ and $\hat{\mu}=1=\gamma$, without any further restrictions on $\aT$, which then takes on the form
\begin{equation}
\aT = 2 f(\phi) \sqrt{X}    \, ,
\end{equation}
where we set $c=\frac{1}{2}$.
In other words, as far as the linearised equations of motion are concerned, theories of the form \eqref{aT_theory} recover the expected GR dynamics of test masses around local sources, without the need for any screening mechanisms. 

An interesting side note is that for the choice $f(\phi) = const$ the theory \eqref{aT_theory} becomes shift-symmetric, with $G_4 \supset \sqrt{X}$. This operator can give rise to hairy black-hole solutions \cite{Babichev:2017guv,Creminelli:2020lxn}, and its presence opens up the possibility of deriving novel constraints on this subclass from the analysis of black-hole physics \cite{Noller:2019chl}. Notice that the resulting non-analytic behaviour as one approaches the Lorentz-invariant vacuum $X = 0$ discussed in \cite{Creminelli:2020lxn} then becomes a worrying feature of these theories, seemingly disconnecting their cosmological dynamics from any Minkowski limit.
\\

\begin{figure}[t!]
\includegraphics[width=\linewidth]{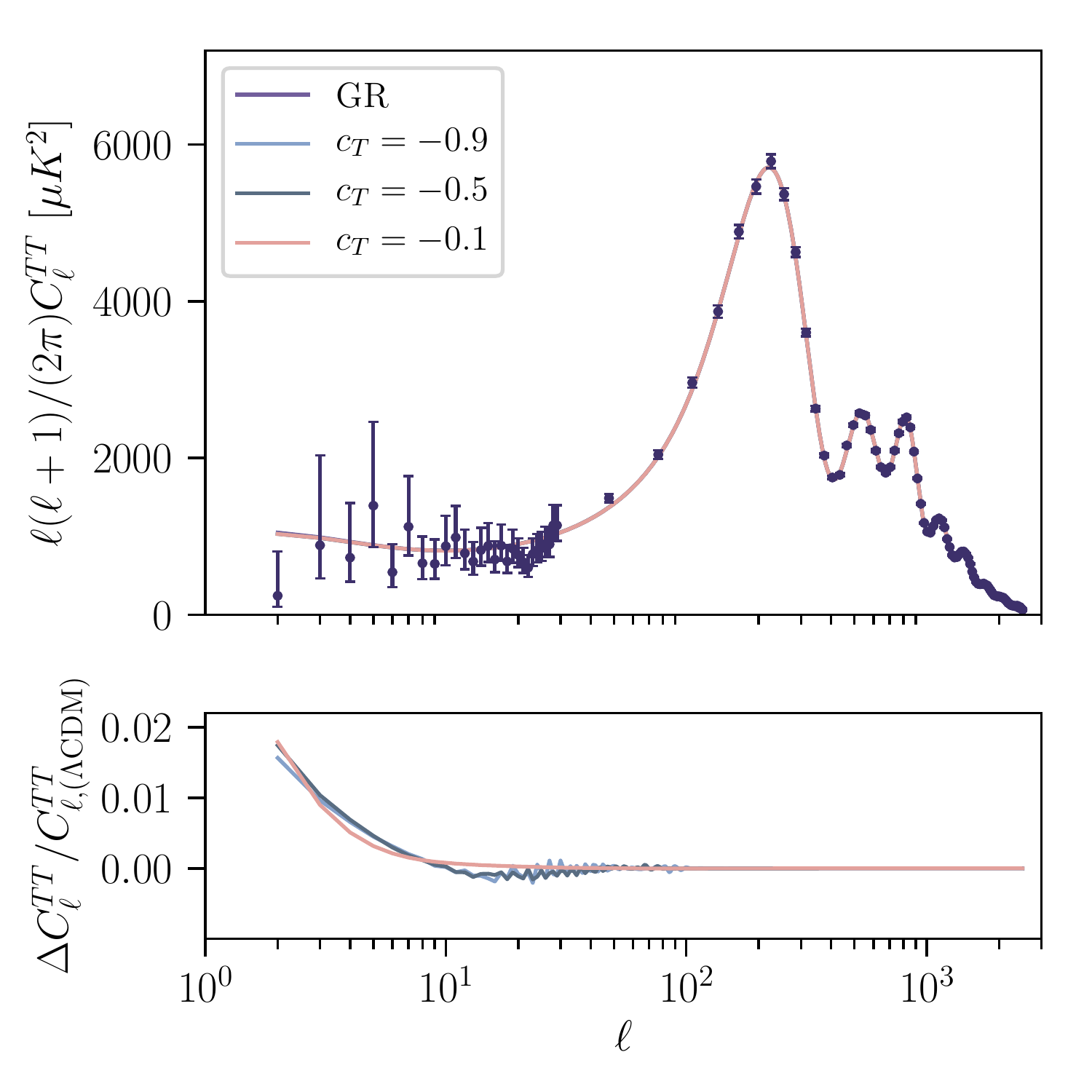} 
    \caption{
Illustration of the effect of varying $c_T$ on the CMB TT power spectrum, where we assume $\aT = c_T \cdot \Omega_{\rm DE}$ and for the reduced Horndeski theory \eqref{aT_theory} with $\aB = 0 = \aM$.
Data points are shown with $1\sigma$ uncertainties and all standard $\Lambda{}$CDM parameters are fixed to their Planck 2015 best-fit values here \cite{Planck-Collaboration:2016ae}. 
As expected from the above consideration of the quasi-static limit, varying $\aT$ only has a negligible effect on cosmological observables here, with small (up to $2\%$) differences to the $\Lambda{}$CDM prediction only appearing on the very largest scales, where they are hidden by cosmic variance -- see the lower panel, where $\Delta {\cal C}_\ell^{TT} = {\cal C}_\ell^{TT} - {\cal C}_{\ell, (\Lambda{\rm CDM})}^{TT}$.
    }%
    \label{fig:cosmo}%
\end{figure}

\nin{\bf Cosmological constraints}: One may now hope to constrain the residual free $\aT$ function with cosmological probes, but since we can see above that $\aT$ drops out in the quasi-static approximation for \eqref{aT_theory} and this approximation is expected to apply to the vast majority of cosmological scales, \eqref{aT_theory} should closely resemble $\Lambda{}$CDM predictions for cosmological observables as well. This is indeed the case and shown in Fig.~\ref{fig:cosmo}, with the only significant deviations from $\Lambda{}$CDM cosmologies arising on very large scales, where the quasi-static approximation starts to break down. 

The main constraint on the residual free $\aT$ function in this setup instead comes from requiring the absence of gradient instabilities, which would e.g.~lead to an unacceptable growth of power in the ${\cal C}_\ell$ power spectrum. 
For a general Horndeski model, requiring the absence of such instabilities amounts to \cite{Kobayashi:2011nu,Bellini:2014fua}
\begin{align} \label{gradient_condition}
(2-\aB)\Big(\bar\alpha - \frac{\dot H}{H^2}\Big) - \frac{3(\rho_{\rm tot} + p_{\rm tot})}{H^2M^2} + \frac{\dot \aB}{H} > 0,
\end{align}
where $\bar\alpha \equiv \tfrac{1}{2}\aB(1+\aT) + \aM - \aT$, we have implicitly assumed the absence of ghost instabilities (requiring $\aK + \tfrac{}{}\aB^2 > 0$) and $\rho_{\rm tot}$ and $p_{\rm tot}$ are the total energy density and pressure in the universe.
For the $\aB = 0 = \aM$ setup considered here, this reduces to just $\aT \leq 0$, where we have assumed a $\Lambda{}$CDM background.
Similarly, from the definition of $\aT$, we know that $\aT < -1$ is associated with an imaginary speed of propagation for the tensor modes (i.e.~gradient instabilities for GWs). Taken together we therefore have
\begin{align}
-1 \leq \aT \leq 0,
\end{align}
where the lower/upper bound comes from requiring the absence of gradient instabilities for GWs/scalar (dark energy) fluctuations. Finally, note that we have of course not applied any bounds on $\aT$ from GW170817 in this section so far.
In essence, they are a measurement of the speed of gravitational waves at energy scales $\sim \Lambda_3$, i.e.~at energy scales somewhat larger than those corresponding to the constraints we have used in this paper otherwise. 
In that sense the bounds discussed in this section are conservative, but if these bounds are deemed to be applicable as well, then we would effectively remove the residual $\aT$-dependence alltogether and again reduce to a pure $G_2$ theory for cosmological purposes.

Finally, note that there are also other observational constraints on $\aT$ at energy scales below $\Lambda_3$, in particular from the Hulse-Taylor pulsar constraining $\aT$ at the $10^{-2}$ level \cite{Jimenez:2015bwa}. We leave an exploration of how precisely to combine this constraint with the others considered here for future work, but taken at face value this constraint is sufficient to impose 
\begin{align}
|\aT| \lesssim 10^{-2},
\end{align}
also in the case when $\aM = 0 = \aB$.

\section{Conclusions} \label{sec:conc}

In this paper we have explored which Horndeski scalar-tensor theories do not require a (Vainshtein) screening mechanism for consistency with fifth force constraints. 
These are the theories for which cosmological backgrounds do not induce any sizeable coupling between matter and scalar fluctuations, so that the scalar does not get sourced by matter and hence cannot mediate any fifth forces in conflict with observations. 
We find that, when constraints for the absence of such a matter-scalar coupling are combined with local solar system bounds, all the $\alpha_i$ parameters controlling deviations from GR around cosmological backgrounds are suppressed by $\gtrsim {\cal O}(10^3)$, i.e.~beyond the reach of near-future cosmological observations. For such cosmological purposes these theories therefore reduce to k-essence theories, although we stress that the higher-order operators associated to the suppressed $\alpha_i$ can in principle still lead to ${\cal O}(1)$ deviations from GR around black holes \cite{Noller:2019chl}.

Closely related to the above, we also investigated which subsets of Horndeski theories precisely recover linearised GR predictions, regardless of whether screening is or is not active on non-linear scales. These are the scalar-tensor theories that survive, should future constraints from linear cosmology eventually become so tight, that no sizeable deviation from GR is permitted on the associated large scales anymore. In terms of the $\alpha_i$, we find that such theories are characterised by $\aM = 0 = \aB$, so the speed of propagation for gravitational waves as measured by $\aT$ is the only discernible difference from linearised GR.
By construction, this $\aT$-dependence drops out of the cosmological observables in the quasi-static limit, so e.g.~has a negligible effect on associated CMB observables.
In this sense, these theories are effectively degenerate with $\Lambda{}$CDM in the linearised (cosmological) regime, even if they originate from a very different covariant Lagrangian---see Eq.~\eqref{aT_theory}---with potentially interesting different behaviours at background and non-linear level. This holds as long as one ignores the GW170817 bound, based on the caveat discussed in \cite{deRham:2018red}. Taking this bound into account instead, rules out the class of theories in \eqref{aT_theory} completely, reducing the admissible operators in \eqref{eq:lagrangian} to $G_2$ only. At a milder level, the same conclusion also holds when taking into account bounds on $\aT$ from binary pulsars \cite{Jimenez:2015bwa}.

The presence of a successful screening mechanism thus indeed appears to be a necessary ingredient to allow for a cosmologically significant presence of the higher-order $G_3$, $G_4$ and $G_5$.
Note that our conclusions are based on the requirement that the absence of deviations from GR is robust under deformations of the scalar profile, although we have never used its explicit form anywhere. Some of the constraints on the $G_i$'s might therefore be relaxed by only requiring them to be satisfied `on-shell', i.e.~upon using the scalar equation of motion (see \cite{Copeland:2018yuh,Bordin:2020fww} for a study along this line and for a discussion of possible obstructions for this approach). The analysis of such a possibility, as well as the generalization to more general scalar-tensor theories \cite{Langlois:2017mdk}, is left for future work.
Finally also note that, even in the presence of screening, current bounds already suggest that the $\alpha_i$ are constrained at the ${\cal O}(10^{-1})$ level \cite{Noller:2020afd} (at least when taking into account bounds from gravitational-wave-induced dark energy instabilities \cite{Creminelli:2019kjy}). While cosmology has long been the most promising arena for detecting signs of new gravitational \dofs{} associated to dark energy, this focus might therefore soon shift to smaller scales, such as the strong field regime probed by gravitational wave observations.
\\

\acknowledgments
We thank Marco Crisostomi, Federico Piazza and Filippo Vernizzi for discussions. JN is supported by an STFC Ernest Rutherford Fellowship, grant reference ST/S004572/1, and also acknowledges support from Dr.~Max R\"ossler, the Walter Haefner Foundation and the ETH Zurich Foundation. 
LS is supported by Simons Foundation Award No.~555117. ET and LGT are supported in part by the MIUR under the contract 2017FMJFMW.
In deriving the results of this paper, we have used: CLASS \cite{Lesgourgues:2011re,Blas:2011rf}, hi\_class \cite{Zumalacarregui:2016pph,Bellini:2019syt} and xAct \cite{xAct}.
\\

\appendix
\section{Quadratic action in Newtonian gauge}
\label{app:qa}

In order to derive the effective Poisson equation \eqref{mrs}, we start from the action \eqref{eq:lagrangian} in the Newtonian gauge \eqref{Def4Pert}  and expand at quadratic order in perturbations. Keeping only the terms with  spatial derivatives acting on the  fields, which are those that are relevant in the quasi-static approximation,  we find  
\begin{align}
 S^{(2)} = & \int \D^4 x \,  \bigg\{  M^2 a \bigg[  
 - 2 \vec{\nabla} \Psi \cdot \vec{\nabla} \Phi    
 + \left(1+\alpha_{\textrm{T}}\right)(\vec{\nabla}\Psi)^2   \nonumber \\
 & + 2\left(\alpha_{\textrm{M}}-\alpha_{\textrm{T}}\right)H   \vec{\nabla} \Psi  \cdot \vec{\nabla}  v_{X}
   + \alpha_{\textrm{B}} H  \vec{\nabla} \Phi   \cdot  \vec{\nabla} v_{X}  
 \nonumber \\
 & +\frac{1}{2}\bigg(2\dot{H}+  \frac{3(\rho_{\rm tot} + p_{\rm tot})}{M^2}  -2\left(\alpha_{\textrm{M}}-\alpha_{\textrm{T}}\right)H^{2}
  \nonumber \\
 & - \frac{\left(aH\alpha_{\textrm{B}}M^2\right)^{.}}{aM^2}  \bigg) (\vec{\nabla} v_{X})^2 
 \bigg]
 + \mathcal{L}_\text{m} \bigg\}
\, ,
\label{STperts-2}
\end{align}
where we introduced $v_{X}\equiv -\delta\phi/\dot{\bar{\phi}}$ and where $\mathcal{L}_\text{m}$ is the Lagrangian for the matter fields. In particular, in the main text, we have considered the case in which $\mathcal{L}_\text{m}=-a^3 \Phi \delta\rho_\text{m}$. The $\alpha_i$-parameters in \eqref{STperts-2} have been previously defined in Eq.~\eqref{alphadef}.
Eq.~\eqref{QSA_scalar} has been obtained from \eqref{STperts-2} after integrating out the gravitational potentials $\Phi $ and $\Psi$. On the other hand, the effective Poisson equation  \eqref{mrs} can be derived from the  equation of motion for $\Phi$ after the fields $\delta\phi$ and $\Psi$ are integrated out.

\bibliographystyle{apsrev4-1}
\bibliography{nsc}

\end{document}